\documentclass[twocolumn, twocolappendix]{aastex63}
\hypersetup{
    linkcolor=[rgb]{0.6,0.3,0.2},
    citecolor=[rgb]{0.1,0.38,0.95},
    filecolor=cyan,
}
\usepackage{amsmath}
\usepackage[utf8]{inputenc}
\usepackage{ulem}
\usepackage{xspace}

\usepackage[encapsulated]{CJK}
\usepackage{ucs}
\newcommand{\cntext}[1]{\begin{CJK}{UTF8}{gbsn}#1\end{CJK}}

\newcommand{\diff}{\ensuremath{\mathrm{d}}}

\newcommand{\Ncluster}{370\xspace}
\newcommand{\Nparam}{372\xspace}
\newcommand{\planck}{\textit{Planck}\xspace}

\received{\today}
\revised{\today}
\accepted{\today}

\submitjournal{ApJ}
\shorttitle{CMB temperature evolution}
\shortauthors{Li et al.}

\begin{document}
\title{Constraining CMB temperature evolution with Sunyaev-Zel'dovich galaxy clusters from the Atacama Cosmology Telescope}
\author[0000-0002-4820-1122]{Yunyang Li (\cntext{李云炀}\!\!)}
\email{yunyangl@jhu.edu}
\affiliation{Department of Physics and Astronomy, Johns Hopkins University, 3701 San Martin Drive, Baltimore, MD 21218, USA}

\author[0000-0003-1690-6678]{Adam D. Hincks}
\affiliation{David A. Dunlap Department of Astronomy \& Astrophysics, University of Toronto, 50 St. George St., Toronto, ON M5S 3H4, Canada}

\author[0000-0002-4200-9965]{Stefania Amodeo}
\affiliation{Department of Astronomy, Cornell University, Ithaca, NY 14853, USA}

\author[0000-0001-5210-7625]{Elia S. Battistelli} \affiliation{Sapienza—University of Rome—Physics department, Piazzale Aldo Moro 5—I-00185, Rome, Italy}

\author[0000-0003-2358-9949]{J. Richard Bond}
\affiliation{Canadian Institute for Theoretical Astrophysics, University of Toronto, 60 St. George St., Toronto, ON M5S 3H8, Canada}

\author{Erminia Calabrese}
\affiliation{School of Physics and Astronomy, Cardiff University, The Parade, Cardiff, Wales, CF24 3AA, UK}

\author{Mark J. Devlin}
\affiliation{Department of Physics and Astronomy, University of Pennsylvania, 209 South 33rd Street, Philadelphia, PA 19104, USA}

\author[0000-0002-7450-2586]{Jo Dunkley}
\affiliation{Joseph Henry Laboratories of Physics, Jadwin Hall, Princeton University, Princeton, NJ 08544, USA}
\affiliation{Department of Astrophysical Sciences, Princeton University, Peyton Hall, Princeton, NJ 08544, USA}

\author[0000-0003-4992-7854]{Simone~Ferraro}
\affiliation{Lawrence Berkeley National Laboratory, One Cyclotron Road, Berkeley, CA 94720, USA}
\affiliation{Berkeley Center for Cosmological Physics, UC Berkeley, CA 94720, USA}

\author{Vera Gluscevic}
\affiliation{Department of Physics and Astronomy, University of Southern California, Los Angeles, CA 90089, USA}

\author[0000-0002-1697-3080]{Yilun~Guan}
\affiliation{Department of Physics and Astronomy, University of Pittsburgh, Pittsburgh, PA 15260, USA}

\author{Mark Halpern}
\affiliation{Department of Physics and Astronomy, University of British Columbia, Vancouver, BC, V6T 1Z4,
Canada}

\author[0000-0002-8490-8117]{Matt Hilton}
\affiliation{Astrophysics Research Centre, University of KwaZulu-Natal, Westville Campus, Durban 4041, South Africa}
\affiliation{School of Mathematics, Statistics \& Computer Science, University of KwaZulu-Natal, Westville Campus, Durban 4041, South Africa}

\author[0000-0002-0965-7864]{Renee Hlozek}
\affiliation{David A. Dunlap Department of Astronomy \& Astrophysics, University of Toronto, 50 St. George St., Toronto, ON M5S 3H4, Canada}
\affiliation{Dunlap Institute for Astronomy and Astrophysics, University of Toronto, 50 St. George St., Toronto, ON M5S 3H4, Canada}

\author[0000-0003-4496-6520]{Tobias A. Marriage}
\affiliation{Department of Physics and Astronomy, Johns Hopkins University, 3701 San Martin Drive, Baltimore, MD 21218, USA}

\author{Jeff McMahon}
\affiliation{Kavli Institute for Cosmological Physics, University of Chicago, Chicago, IL 60637, USA}
\affiliation{Department of Astronomy and Astrophysics, University of Chicago, 5640 S. Ellis Ave., Chicago, IL 60637, USA}
\affiliation{Department of Physics, University of Chicago, Chicago, IL 60637, USA}
\affiliation{Enrico Fermi Institute, University of Chicago, Chicago, IL 60637, USA}
\affiliation{Department of Physics, University of Michigan, Ann Arbor, MI 48109, USA}

\author{Kavilan Moodley}
\affiliation{Astrophysics Research Centre, University of KwaZulu-Natal, Westville Campus, Durban 4041, South Africa}
\affiliation{School of Mathematics, Statistics \& Computer Science, University of KwaZulu-Natal, Westville Campus, Durban 4041, South Africa}

\author[0000-0002-4478-7111]{Sigurd Naess}
\affiliation{Center for Computational Astrophysics, Flatiron Institute, New York, NY 10010, USA}

\author[0000-0002-8307-5088]{Federico Nati}
\affiliation{Department of Physics, University of Milano-Bicocca, Piazza della Scienza 3, 20126 Milano (MI), Italy}

\author{Michael D. Niemack}
\affiliation{Department of Physics, Cornell University, Ithaca, NY 14853, USA}
\affiliation{Department of Astronomy, Cornell University, Ithaca, NY 14853, USA}

\author[0000-0003-1842-8104]{John Orlowski-Scherer}, 
\affiliation{Department of Physics and Astronomy, University of Pennsylvania, 209 South 33rd Street, Philadelphia, PA 19104, USA}

\author[0000-0002-9828-3525]{Lyman Page}
\affiliation{Joseph Henry Laboratories of Physics, Jadwin Hall, Princeton University, Princeton, NJ 08544, USA}

\author[0000-0001-6541-9265]{Bruce Partridge}
\affiliation{Department of Physics and Astronomy, Haverford College, Haverford, PA 19041, USA}

\author{Maria Salatino}
\affiliation{Physics Department Stanford University, Stanford, CA 94305, USA}
\affiliation{Kavli Institute for Particle Astrophysics and Cosmology (KIPAC), Stanford, CA 94305, USA}

\author[0000-0002-4619-8927]{Emmanuel Schaan}
\affiliation{Lawrence Berkeley National Laboratory, One Cyclotron Road, Berkeley, CA 94720, USA}
\affiliation{Berkeley Center for Cosmological Physics, UC Berkeley, CA 94720, USA}

\author{Alessandro Schillaci}
\affiliation{Department of Physics, California Institute of Technology, Pasadena, CA 91125, USA}

\author{Neelima Sehgal}
\affiliation{Physics and Astronomy Department, Stony Brook University, Stony Brook, NY 11794, USA}

\author[0000-0002-8149-1352]{Crist\'obal Sif\'on}
\affiliation{Instituto de F\'isica, Pontificia Universidad Cat\'olica de Valpara\'iso, Casilla 4059, Valpara\'iso, Chile}

\author{Suzanne T. Staggs}
\affiliation{Joseph Henry Laboratories of Physics, Jadwin Hall, Princeton University, Princeton, NJ 08544, USA}

\author{Alexander\ van Engelen}
\affil{School of Earth and Space Exploration, Arizona State University, Tempe, AZ 85287, USA}

\author[0000-0002-7567-4451]{Edward J. Wollack}
\affiliation{NASA/Goddard Space Flight Center, Greenbelt, MD 20771, USA}

\author[0000-0001-5112-2567]{Zhilei Xu}
\affiliation{Department of Physics and Astronomy, University of Pennsylvania, 209 South 33rd Street, Philadelphia, PA 19104, USA}
\affiliation{MIT Kavli Institute, Massachusetts Institute of Technology, 77 Massachusetts Avenue, McNair Building, Cambridge, MA 02139, USA}

\begin{abstract}
The Sunyaev-Zel'dovich (SZ) effect introduces a specific distortion of the blackbody spectrum of the cosmic microwave background (CMB) radiation when it scatters off hot gas in clusters of galaxies.
The frequency dependence of the distortion is only independent of the cluster redshift when the evolution of the CMB radiation is adiabatic.
Using \Ncluster clusters within the redshift range $0.07\lesssim z\lesssim1.4$ from the largest SZ-selected cluster sample to date from the Atacama Cosmology Telescope,
we provide new constraints on the deviation of CMB temperature evolution from the standard model $\alpha=0.017^{+0.029}_{-0.032}$, where $T(z)=T_0(1+z)^{1-\alpha}$.
This result is consistent with no deviation from the standard adiabatic model. 
Combining it with previous, independent datasets we obtain a joint constraint of $\alpha=-0.001\pm0.012$.
\end{abstract}

\keywords{
    \href{http://astrothesaurus.org/uat/343}{Cosmology (343)};
    \href{http://astrothesaurus.org/uat/322}{Cosmic microwave background radiation (322)};
    \href{http://astrothesaurus.org/uat/584}{Galaxy clusters (584)};
    \href{http://astrothesaurus.org/uat/1654}{Sunyaev-Zeldovich effect(1654)};
}

\section{Introduction} \label{sec:intro}
The cosmic microwave background (CMB) is an almost-perfect blackbody with a temperature today of $T_0 = (2.72548 \pm 0.00057)$\,K \citep{FIRAS_T}. In the standard cosmological model, this radiation fills a universe characterized by a Friedmann--Lema\^{i}tre--Robertson--Walker (FLRW) metric,
\begin{equation}
  \diff s^2 = -c^2 \diff t^2 + a^2(t)\diff\mathbf{x}^2,
\end{equation}
where $a = (1+z)^{-1}$ is the scale factor that describes the proper size of the spatial component $\mathbf{x}$. Under the assumption of adiabatic expansion---or, equivalently, conservation of the energy-momentum tensor---one can show that the radiation energy density $u \propto T^4$ scales as $u \propto a^{-4}$. Therefore, the associated radiation temperature must evolve as $T(z) = T_0 (1 + z)$. Empirically probing this temperature--redshift relation is thus a test of some of the most fundamental assumptions in cosmology. Deviations would indicate either that the FLRW metric does not describe our Universe or else that the CMB does not behave adiabatically. 
In the first case, the cosmological principle of homogeneity and isotropy would be violated. Though isotropy is well-established observationally, we could conceivably be at the centre of a large scale, isotropic inhomogeneity, such as a void \citep{goodman:1995, Clarkson2012}. 
In the other case of non-adiabaticity, energy would be injected into or removed from the CMB by exotic physics such as vacuum energy decay \citep{Lima1996,Lima+2000, Jetzer2011}. Finally, variation of fundamental constants such as the fine-structure constant can mimic non-adiabatic behavior in the observables from which $T(z)$ is reconstructed \citep[e.g.,][]{demartino/etal:2016}.

It is customary to parameterize deviations from the expected evolution as  \citep{Lima+2000}:
\begin{equation}
    T(z) = T_0 (1+z)^{1-\alpha}.
    \label{eq:def_alpha}
\end{equation}
Non-zero $\alpha$ indicates deviation from the standard cosmology. The parameterization of Equation~\ref{eq:def_alpha} is motivated by the scenario in which the increase (or decrease) in photon number is `adiabatic', i.e., in which the specific entropy of the CMB radiation remains constant \citep{Lima+2000}. However, since $\alpha$ is small, a Taylor expansion of Equation~\ref{eq:def_alpha} can be used to assess other models in the $z < \mathcal{O}(1)$ regime \citep{Avgoustidis2016}.  \citet{Chluba2014} observes that unless the energy spectrum of the injected/removed photons has a very particular form, this mechanism introduces distortions into the CMB blackbody spectrum that are already strongly constrained by the Cosmic Background Explorer (\textit{COBE}) Far InfraRed Absolute Spectrophotometer (FIRAS; \citealt{FIRAS_T}). Nevertheless, probing $\alpha$ still provides a valuable empirical check of the validity of our cosmological model. It not only serves as a `null test' against the finely-tuned case in which adiabatic energy injection/removal causes no spectral distortions, but is also in principle sensitive to large scale inhomogeneity in the metric \citep[see][]{Chluba2014}.

Two methods have been used to directly probe $T(z)$. 
The first uses line spectroscopy of quasars to identify, for instance, the fine-structure line of \ion{C}{1} or molecular rotational transitions of CO due to absorption of CMB radiation. \cite{Avgoustidis2016} combined the result from 10 quasar absorption line systems at $z = 0.89{-}3.025$ \citep{Srianand2000, Ge2001,Molaro2002, Cui2005,  Srianand2008, Noterdaeme2010, Noterdaeme2011, Muller2013} and provided a constraint of $\alpha = 0.005 \pm 0.022$. Recently, \citet{klimenko/etal:2021} updated the line modeling by correcting for the collisional excitation of CO rotational transitions and obtained a constraint of $\alpha=-0.015^{+0.030}_{-0.028}$ from 12 systems. 
The second method, which we use in this paper, exploits the fact that the amount of inverse Thomson scattering of CMB photons in galaxy clusters, known as the Sunyaev-Zeldovich (SZ) effect \citep{zeldovich/sunyaev:1969, sunyaev/zeldovich:1972}, depends on $T(z)$. The scattering distorts the CMB blackbody spectrum to induce a frequency-dependent change in its intensity, or (as is more convenient to work with) its thermodynamic temperature:
\begin{equation}
\Delta T(\hat{r}, \nu) = T_0 \int \mathrm{d}r\,n_{\mathrm{e}}(\hat{r}, z) \sigma_{\mathrm{T}} \left[f_\mathrm{SZ}(\nu, T_{\mathrm{e}})\frac{k_\mathrm{B} T_{\mathrm{e}}}{m_{\mathrm{e}}c^2} - \frac{v_r}{c}\right],
\end{equation}
where the integral is done along the line of sight in direction $\hat{r}$, $n_{\mathrm{e}}$ is the number density of electrons, $T_{\mathrm{e}}$ is their temperature, $v_r/c$ is the proper radial velocity of the cluster relative to the speed of light, $k_\mathrm{B}$ is the Boltzmann constant, $m_{\mathrm{e}}$ the mass of the electron, and $\sigma_{\mathrm{T}}$ is the Thomson cross section. The function $f_\mathrm{SZ}(\nu, T_{\mathrm{e}})$ encodes the frequency dependence of the SZ effect. Normally, one makes the approximation that $T_{\mathrm{e}}$ is constant through the cluster to simplify the expression to:
\begin{equation}
  \Delta T(\hat{r}) = T_0 \left[f(\nu, T_{\mathrm{e}})y_\mathrm{c}(\hat{r})-\tau\frac{v}{c}\right],
  \label{eq:sz_with_y}
\end{equation}
where $y_c$ is the Compton parameter, proportional to the integrated gas pressure, $n_{\mathrm{e}} T_{\mathrm{e}}$, and $\tau$ is the integrated optical depth along the line of sight. The last term in Equation~\ref{eq:sz_with_y} is the kinematic SZ effect (kSZ) and has the same spectral shape as the CMB, whereas the former characterizes the thermal SZ (tSZ) effect and has the spectral shape
\begin{equation}
f(\nu)=x\coth(x/2)-4, \quad x\equiv\frac{h\nu(1+z)}{k_\mathrm{B}T(z)}, 
\end{equation}
in the non-relativistic approximation of small $T_{\mathrm{e}}$,\footnote{We investigate the effect of the full relativistic treatment on our result in \S\ref{sec: result}.} where $h$ is the Planck constant. If $\alpha = 0$, then the $1+z$ in the numerator is cancelled out (see Equation~\ref{eq:def_alpha}) and $f(\nu)$ is independent of redshift. If, on the other hand, the temperature deviates from the canonical form, it can be probed by multi-frequency observations \citep{rephaeli:1980}. For instance, the ratio of the SZ amplitude measured at two different frequencies,  $\nu_1$ and $\nu_2$, will depend on the redshift:
\begin{equation}
r(z) =
  \frac{f(\nu_1, z)}{f(\nu_2, z)}.
  \label{eq:sz_ratio}
\end{equation}
Measuring $r(z)$ over a range of redshifts can thus constrain $\alpha$.

This technique was first used by \citet{Battistelli2002} using two clusters (Coma and Abell~2163) observed with multiple observatories at four frequencies between 30 and 270\,GHz to measure an $\alpha$ consistent with zero with a $2\sigma$ uncertainty of $\Delta\alpha \sim 0.3$. Since then, results using increasing numbers of clusters have found no evidence for non-zero $\alpha$ with shrinking uncertainties (all $1\sigma$): \citet{Luzzi2009} analysed nine clusters using six frequencies between 30 and 353\,GHz from multiple observatories and found $\Delta\alpha \sim$ 0.06--0.09; \citet{SPT2014} used 158 clusters observed by the South Pole Telescope at 95 and 150\,GHz to measure $\Delta\alpha = 0.03$; \citet{Hurier+2014} used maps from the \planck satellite to analyze 813 clusters from the first catalog of \planck clusters (PSZ1) and reported $\Delta\alpha = 0.017$, while \citet{Luzzi2015} used a different pipeline on the same maps and 103 clusters from PSZ1, yielding $\Delta\alpha = 0.016$; \citet{Planck2015_martino}, on the other hand, used 481 X-ray selected clusters from \textit{ROSAT} to analyze the \planck data and measured $\Delta\alpha = 0.013$, but with a possible systematic of up to $\Delta\alpha = 0.02$ (with the sign being the same as that of $\alpha$) coming from their map-cleaning process. Combining all the foregoing quasar and SZ data, a constraint of $\Delta\alpha = 0.013$ is achieved \citep{Avgoustidis2016, klimenko/etal:2021}.\footnote{This is slightly larger than the corresponding value reported in \citep{Avgoustidis2016}, as we include the systematic error reported by \citet{Planck2015_martino}.}

In this paper, we provide an updated measurement of $\alpha$ with the largest catalog of SZ clusters from the Atacama Cosmology Telescope (ACT) out to a redshift of 1.4 (Section~\ref{sec: data}). 
This work improves on the likelihood analysis used in previous studies (Section~\ref{sec:parameter_inference}) and validates it with simulations (Section~\ref{sec: sim}). We also examined the potential systematics in the result, some of which have been neglected in similar analyses (Section~\ref{sec: result}).

\section{Data} \label{sec: data}
We use the ACT DR5 maps \citep{ACT_DR5_coadd} at 98\,GHz and 150\,GHz for this analysis.\footnote{DR5 maps and ancilliary products are available at: \url{https://lambda.gsfc.nasa.gov/product/act/actpol_prod_table.cfm}}
The data include both nighttime observations from 2008 to 2018 and daytime observations from 2014 to 2018, covering 18,000 deg$^2$ of the sky.
Maps were made by co-adding individual, maximum-likelihood maps from each observation season and detector array. 
The final, convolved maps at 98 and 150\,GHz have FWHM resolutions of $2.2\arcmin$ and $1.4\arcmin$, respectively.

\subsection{SZ Sample and Photometry} \label{sec: matched-filter}
Our SZ cluster sample comes from \cite{Hilton2020}, which is the largest homogeneous SZ-selected catalog to date, containing 4,195 optically confirmed clusters within a search area of 13,211\,deg$^2$. 
The sample selection in \cite{Hilton2020} assumes that the relative amplitude of the two bands follows the adiabatic model ($\alpha=0$). 
We test in Section \ref{sec: pseudo-sim} that this does not bias our result towards a false negative.

SZ signals are measured with a matched-filter technique \citep{Melin2006} in two steps. Cluster shapes and locations are extracted with a multi-band matched filter applied to each map at frequency $\nu_i$:
\begin{equation}
    \psi(\nu_i, \mathbf{k})=\sum_{j}\mathbf{N}_{ij}^{-1}(\mathbf{k})B_{j}(k)f_\mathrm{SZ}(\nu_j)S(k, \nu_j),
\end{equation}
where $S$ is the Universal Pressure Profile (UPP) for galaxy clusters \citep{Arnaud_2010}, $B$ is the beam window function, and $\mathbf{N}_{ij}$ is the noise covariance matrix that includes contributions from instrumental noise and non-tSZ sky signals.
We estimate $\mathbf{N}_{ij}$ as in \cite{Hilton2020}, except that we update this method by doing two passes of filtering.
After filtering the map once, we find all clusters with signal-to-noise ratio (S/N) $>5$, subtract them from the map, and use it as noise estimation for the second pass of filtering.
This multi-band matched-filter is applied for 31 candidate profiles with projected angular size $\theta_{500c}$ (defined as the cluster radius enclosing an average density 500 times the critical density of the Universe at the cluster redshift) log-uniformly spaced from 0.6$\arcmin$ to 8$\arcmin$. The profile $S$ that maximises the S/N for each cluster is then selected for signal extraction.\footnote{We verified that 31 filters is sufficiently precise by performing the analysis on a subset of 12, sparsely-spaced profiles and finding consistent results.}

Next, the SZ signal is measured in each band with spatial matched-filter forced photometry. This has been demonstrated to be robust against Galactic contamination and infrared emission from the galaxy clusters \citep{2018MNRAS.476.3360E}.
We smooth the $150\,\mathrm{GHz}$ map down to the same angular resolution as the $98\,\mathrm{GHz}$ map ($2.2\arcmin$) with point spread function (PSF) matching, and filter both maps with the same Fourier kernel:
\begin{equation}
    \psi_{98, 150}(\mathbf{k})=\mathrm{N}^{-1}_{98}(\mathbf{k})B_{98}(k)S(k),
\end{equation}
where $S$ is the cluster profile determined from the multi-band matched filter as described above.
Although the filter at $150\,\mathrm{GHz}$ is not the optimal matched filter, this technique ensures that the potential bias from a filter-mismatch is a multiplicative factor common to both frequencies that is marginalized over in the likelihood analysis (Section \ref{eq:likelihood}), and thus leaves the constraints on $\alpha$ unbiased \citep{SPT2014}.

\subsection{Sample selection}
\label{ssec:sample_selection}

A S/N threshold of $\xi = 5.5$ for the SZ signal measurement at each frequency is applied to eliminate samples that are more susceptible to systematics, such as primary CMB anisotropies or infrared/radio emission. 
This particular choice of S/N cut is conservative and is informed by  simulations (see Section \ref{sec: websky-sim}).
However, it inevitably biases the signal towards higher values since positive fluctuations will be preferentially included. Because this bias will not necessarily be the same in both frequencies, it may not cancel out in the ratio between the two bands that we use in our analysis, and so we correct for it in our likelihood calculation (Section ~\ref{sec:parameter_inference}).

To eliminate potential contamination from bright radio sources, we exclude clusters found within $2\arcmin$ of sources brighter than 10\,mJy in the NRAO VLA Sky Survey \citep[NVSS,][]{NVSS} 1.4\,GHz catalog, or sources brighter than 15\,mJy in the Sydney University/Molonglo Sky Survey  \citep[SUMSS][]{SUMSS} 843\,MHz catalog. 
The combination of these two catalogs covers the full survey region of ACT DR5. 
Assuming a synchrotron spectral index $-0.7$ in flux (Dicker et al., in prep.), this ensures the bias is below 3\% for 99\% of the sample at 98 and 150 \,GHz. 
Tests in Section~\ref{sec: result} show that our result does not change when a stricter limit is applied.

The two individual selections remove 3627 and 929 clusters respectively, and leave our final sample with \Ncluster clusters in a redshift range of $0.07\lesssim z\lesssim1.4$.

\subsection{Band centers and calibration}\label{sec: calibration}

The effective band center for the tSZ signal is obtained by integrating its spectrum over the detector bandpass.
We ignore the change in the band center due to deviation of the standard tSZ spectrum for models with non-adiabatic temperature evolution, as the difference is negligible.
However, the effective bandpasses can vary across the map due to different combinations of detectors being coadded in different regions. Following the prescription of \citet[][Appendix A.4]{ACT_DR5_coadd}, we calculate the resulting effective bandpass centers in a $0.5^{\circ}\times0.5^{\circ}$ grid,\footnote{The data products required for this are available at: \url{https://lambda.gsfc.nasa.gov/product/act/actpol_dr5_aux_prod_get.cfm}}
finding that the mean band centers for all clusters are 97.4 and 147.8\,GHz, respectively, with a variation that translates to a $1\%$ variation in the tSZ signal ratio between the two frequencies. 
We verified that this variation of band centers across the map does not change our final result.
Due to uncertainties in the detector bandpass measurement, there is a $\sim1.5/2.4\,\mathrm{GHz}$ systematic shift in the band center estimate at $98/150$ GHz respectively \citep{Madhavacheril2020}. This results in a $2\%$ uncertainty in the tSZ signal ratio.
Furthermore, the ACT DR5 maps are separately calibrated against \planck to the $\mathcal{O}(1\%)$ level \citep{ACT_DR5_coadd}. 
Assuming these uncertainties are independent at the two frequencies, the relative calibration uncertainty is $\mathcal{O}(1.5\%)$.
Combining the two, we adopt a $2.5\%$ prior on the relative calibration between our two frequency channels to account for the instrumental uncertainties.

\section{Parameter Inference}
\label{sec:parameter_inference}
Given the temperature evolution described in Equation ~\ref{eq:def_alpha}, assuming Gaussian errors, and accounting for the S/N threshold $\xi$, the likelihood function is \citep[e.g.,][\S4.2.7]{Ivezic2014}:
\begin{widetext}
\begin{eqnarray}
\mathcal{L}_{i}(s_{i,98}, s_{i,150}|\alpha,\eta,\mu_{i,150}) = \frac{4}{2\pi\sigma_{i,98}\sigma_{i, 150}} \frac{\exp{\left[-\frac{(s_{i,98}-\eta r(\alpha, z_i)\mu_{i,150})^2}{2\sigma_{i,98}^2}\right]}}{\mathrm{erfc}\left[\frac{\xi\sigma_{i, 98}-\eta r(\alpha, z_i)\mu_{i, 150}}{\sqrt{2}\sigma_{i, 98}}\right]}\frac{\exp{\left[-\frac{(s_{i,150}-\mu_{i, 150})^2}{2\sigma_{i,150}^2}\right]}}{\mathrm{erfc}\left[\frac{\xi\sigma_{i, 150}-\mu_{i, 150}}{\sqrt{2}\sigma_{i, 150}}\right]},
\label{eq:likelihood}
\end{eqnarray}
\end{widetext}
where $s_i$ are the measured tSZ signals, and the uncertainties $\sigma$ are taken as known quantities, which are taken to be independent between two frequencies for each cluster. 
We note that this is a simplification since the primary CMB anisotropy and astrophysical emission introduce correlated errors in the tSZ signal estimation even though the instrumental noise could be treated as independent. 
This correlation is accounted for in Section \ref{sec: result} and is found to be insignificant in altering the result.
We parameterize the fitting model with the amplitude of the tSZ signal at 150\,GHz, $\mu_{i, 150}$, and the ratio between the signals in the two bands, $r(\alpha, z_i)\equiv\mu_{i, 98}/\mu_{i, 150}$ (Equation~\ref{eq:sz_ratio}). 
The parameter $\eta$ accounts for any relative calibration difference between the two bands \citep[c.f.,][]{Luzzi2009} and is marginalized over in our results.
As discussed in Section \ref{sec: calibration}, we use a Gaussian prior with $2.5\%$ standard deviation for $\eta$.
Finally, the probability distribution functions of $s_{i, 98}$ and $s_{i, 150}$ are truncated to 0 below the S/N threshold $\xi$, and the  complementary error functions in the denominator account for their normalization. 
When the value of $\mu_{i,150}/\sigma_{i, 150}$ (or, for $98$ GHz, $\eta r\mu_{i,150}/\sigma_{i, 98}$) is well above the threshold (e.g., $\gtrsim \xi + 2$), the complementary error function approaches an asymptotic value of 2 and the likelihood reduces to the standard Gaussian form.
We note that \cite{SPT2014} and \cite{ Planck2015_martino} do not account for this effect, which could in principle bias $\alpha$, but would depend on the S/N distribution of their samples.

The free parameters in the model are $\{\alpha, \eta, \mu_{i, 150}\}$. For $\Ncluster$ clusters, the total number of parameters is thus $\Nparam$.
The model amplitudes $\mu_{i,150}$ are essentially nuisance parameters over which we integrate, assuming flat prior distributions, to obtain the marginalized posterior distributions of $\alpha$ and $\eta$:

\begin{align}
\log \mathcal{P}(\alpha, \eta) = \sum_i  \log \int \mathcal{L}_i(s_{i,98}, s_{i,150}|\alpha,\eta,\mu_{i,150})\mathrm{d}\mu_{i, 150}. \label{eq:posterior}
\end{align}
For the final estimation of $\alpha$, we report the mean of the posterior distribution after marginalizing over $\eta$ in Equation \ref{eq:posterior}.
Since the individual likelihood function for each cluster in Equation \ref{eq:likelihood} is non-Gaussian and the high dimensionality of the problem could lead to a volume effect such that the posterior mean might be offset from the maximum {\it a posteriori} (MAP) value, we also report the MAP values for reference.
They are calculated by maximizing 
\begin{align}
\log \mathcal{L}_\mathrm{ML}(\alpha, \eta) = \sum_i  \log \mathcal{L}_i(s_{i,98}, s_{i,150}|\alpha,\eta,\hat{\mu}_{i,150}),\label{eq:MLL}
\end{align}
where $\hat{\mu}(\alpha, \eta)_{i,150}$ is the maximum-likelihood estimate of $\mu_{i, 150}$ for a given $\alpha$, determined by:
\begin{align}
\left.\partial_{\mu_{i, 150}} \mathcal{L}_i(s_{i,98}, s_{i,150}|\alpha,\eta, \mu_{i,150})\right|_{\hat{\mu}_{i,150}} \equiv 0.
\end{align}
When the selection bias correction is ignored, Equation~\ref{eq:MLL} reduces to the form that is used by \citet{SPT2014} as the likelihood function (their Equation~5).
However, we note that this is merely a cross section of the true multivariate likelihood function (Equation \ref{eq:posterior}) that fixes all the nuisance parameters at their maximum-likelihood values.
While its peak value can be used for MAP estimation, the quantile range of this distribution does not include the covariances with $\{\mu_{i, 150}\}$, and is an underestimation of the true uncertainty.
For our MAP estimation, we quote the standard deviation of the marginalized posterior distribution as the $1\sigma$ uncertainty.

We use a Markov-chain Monte Carlo sampler \citep[\texttt{emcee}, ][]{emcee} to find the posterior distribution of Equation~\ref{eq:posterior} and the maximum-likelihood solution to Equation~\ref{eq:MLL} .
To ensure convergence of the MCMC, we compute the maximum autocorrelation of the chains, which we find to be $T = 30$ iterations, and discard the first $7\,T$ iterations from our results \citep[see][]{emcee}.\footnote{We have also run the MCMC for our baseline sample of clusters to over 2,500 iterations and confirmed that this autocorrelation length is stable and that having a longer burn-in period does not alter our results.}

\section{Simulations} \label{sec: sim}
\subsection{\texttt{WebSky} simulation} \label{sec: websky-sim}
To validate the method outlined above, we applied it to simulated ACT observations using inputs from \texttt{WebSky} \citep{Websky}. 
This simulation of the cosmic web at millimeter wavelengths includes the CMB, the cosmic infrared background (CIB) and the SZ effect. 
The AGN feedback effect on the cluster pressure profile is also considered in the simulation with a prescription from \cite{Battaglia2012}.
By mixing different components from the \texttt{WebSky} maps, we are able to simulate two scenarios: first, a sky with only the CMB and the tSZ signals with $\alpha=0$, and second, a sky that also includes the CIB and the kSZ effect.
We add ACT DR5 white noise realizations to the simulations, as realizations of ACT's more complex noise properties \citep{ACT_DR5_coadd} are not available. However, we are looking at small scales where the white noise approximation should suffice for the purposes of testing how the different signal components affect our method.\footnote{Using white noise can overestimate the cluster S/N, as noted by \citep{Hilton2020} who also used this approximation, but we verify that our results are robust against higher S/N cuts (Section~\ref{sec: result}).}
The cluster extraction and photometry pipelines are the same as used in the analysis of real data. 

A positive correlation between the CIB and the cluster tSZ signal is expected \citep{Addison+2012}  and has been observed by various experiments \citep{Dunkley+2013, Gerorge+2015, Planck23}.
As infrared emission from dust-obscured star formation peaks in the redshift range $1\lesssim z \lesssim 2$, its correlation with the tSZ signal is most significant at the higher end of the redshift distribution of our sample \citep[e.g.,][]{Addison+2012,Madau/Dickinson2014}.
For most of the sources close to the center of the matched filter, this contamination acts as a negative bias to the tSZ decrement measurement (but due to the ringing of the filter, the infill could also lead to a positive bias for sources in the intermediate radii from the center; Dicker et al., in prep.), and has more impact on the 150\,GHz band and thus translates into a positive bias in $\alpha$.

\begin{figure}
    \centering
    \epsscale{1.25}
    \plotone{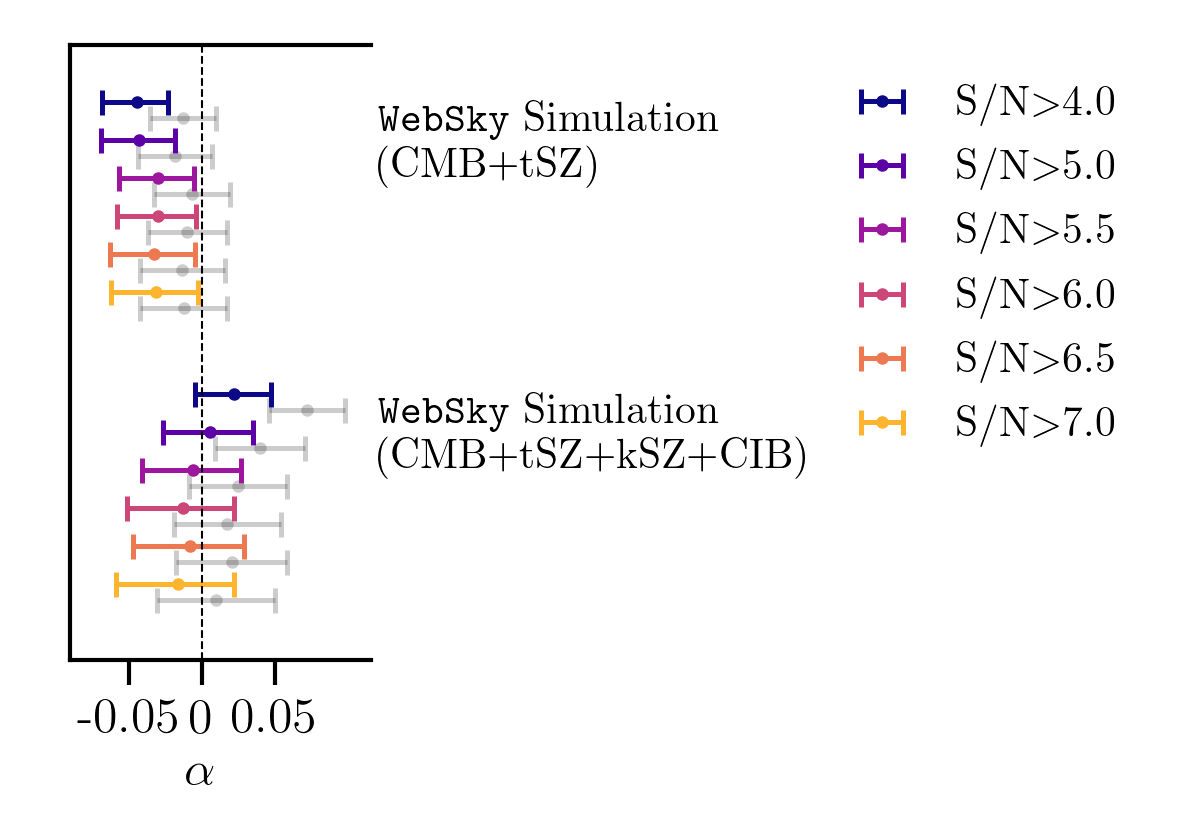}
    \caption{Constraints on $\alpha$ from \texttt{WebSky} simulations with input $\alpha=0$. The colored points indicate the mean of the marginalized posterior distributions with $1\sigma$ error bars. The grey points indicate the MAP value with the same error bars as the colored ones. Based on the CMB+tSZ+kSZ+CIB simulation results, we restrict the analysis to data with S/N~$>5.5$ to avoid contamination from infrared sources. }
    \label{fig:sim_websky}
\end{figure}

We perform the analysis with various S/N cuts $\xi$ in each band, and report in Figure~\ref{fig:sim_websky} the posterior mean (in colors) and MAP (in grey) estimation of $\alpha$ as described in Section \ref{sec:parameter_inference}.
For this realization, posterior mean and MAP estimation are offset by ${\sim}1\sigma$, which could be related to the volume effect.

For the \texttt{WebSky} simulation that includes CMB+tSZ, our result is consistent with $\alpha=0$ for all S/N selection within $68\%$ credible intervals.
This also demonstrates that using a fixed UPP profile for signal estimation is sufficient for our uncertainty, despite the fact that the tSZ profile might have redshift dependency due to feedback.
Although the posterior mean and MAP are offset from each other, their trends are similar as the S/N is varied.
Figure~\ref{fig:sim_websky} indicates that the inclusion of the CIB component may introduce positive bias in $\alpha$ at $\xi \lesssim 5$.
This bias appears to be mitigated by choosing clusters with higher S/N, as those clusters are more tSZ-dominated, massive clusters.
Guided by this test, we adopt $\xi=5.5$ in our analysis.
We verify in Section \ref{sec: result} that our result is insensitive to the choice of S/N threshold beyond $5.5$.

In the validation of our method described in this section we are limited by having only one \texttt{WebSky} realization. 
We did a simple check by reversing the coordinates of the clusters in the map (flipping them both north--south and east--west), thereby changing the S/N distribution of the sample. In this case, the difference between the CMB+tSZ and CMB+tSZ+kSZ+CIB posteriors is little larger (but $\lesssim 1\sigma$), and the positive bias in $\alpha$ at $\xi \lesssim 5$ is less apparent. However, we find broad agreement with the results above; and we retain the $\xi = 5.5$ threshold out of caution for the possible effect of the CIB.

\subsection{Pseudo-simulation}\label{sec: pseudo-sim}
We further validate our pipeline with a suite of pseudo-simulations.
For these simulations, we insert UPP-model clusters into maps containing a CMB
realization and white noise that follows the ACT DR5 inverse variance maps. 
The mock cluster sample is drawn from the \citet{Tinker2008} halo mass function, assuming a modified version of the \citet{Arnaud_2010} SZ-mass--scaling relation, with the normalization adjusted to approximately reproduce the number of clusters that are observed in the real maps.
This test simulates the effect of a non-standard temperature evolution by setting the injected SZ signal at the two observing frequencies according to specified values of $\alpha$. 
As all clusters are selected with a multi-frequency matched-filter assuming the standard adiabatic model \citep{Hilton2020}, this test is crucial to preclude the possibility of a false-negative non-detection of the deviation.

We use three inputs for the temperature evolution, $\alpha=0$, $-0.1$, and $0.05$, and generate ten pseudo-catalogs for each of these inputs.
Similar to \cite{Hilton2020}, we search for clusters using the multi-frequency matched filter assuming $\alpha=0$ with detection threshold $\xi=4$, and perform the forced photometry as outlined above using selection $\xi=5.5$. 
Without applying the prior on $\eta$ and using ACT beam profile for simulation, the posterior mean agrees with $\eta=1$ in all three cases, but the recovered $\eta$ is biased by $0.01$ (at the $3\sigma$ level according to the error on the mean of the 10 simulations) if we use a Gaussian beam profile with similar beam size. 
This is likely due to the imperfect PSF matching we used when smoothing the 150\,GHz maps to the same scale as 98\,GHz maps in the Gaussian case. While we don't find a similar offset when using the ACT PSF, out of an abundance of caution, we choose to augment the $2.5\%$ prior width based on instrument calibration and bandpass (Section \ref{sec: calibration}) by an uncorrelated 1\% uncertainty. 
Using the updated $2.7\%$ prior on $\eta$, we find that the recovered $\alpha$ still agrees with the inputs, with average (over 10 simulations) $\alpha=-0.004\pm0.006$, $-0.102\pm0.007$ and $0.047\pm0.005$, respectively.
This test verifies that our pipeline is not preferentially selecting clusters that appear, by chance, close to $\alpha=0$.
A similar conclusion was also reached by \cite{SPT2014}.

\section{Results} \label{sec: result}
\begin{figure}
    \centering
    \epsscale{1.2}
    \plotone{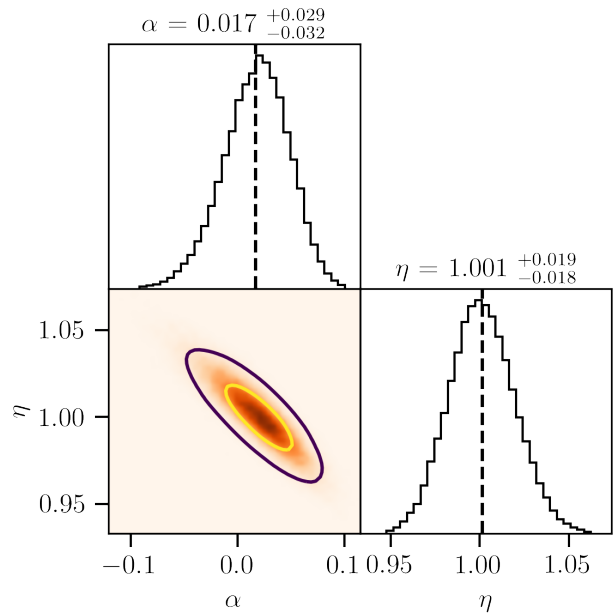}
    \caption{The posterior distribution of $\alpha$ and $\eta$ after marginalizing over nuisance parameters $\{\mu_{i, 150}\}$. 
    The contours mark 68 and 95\% confidence intervals in the marginalized 2D space.
    Reported uncertainties are derived from the 68\% quantile of the marginalized 1D distributions.}
    \label{fig:posterior_corner}
\end{figure}
\begin{figure}
    \centering
    \epsscale{1.2}
    \plotone{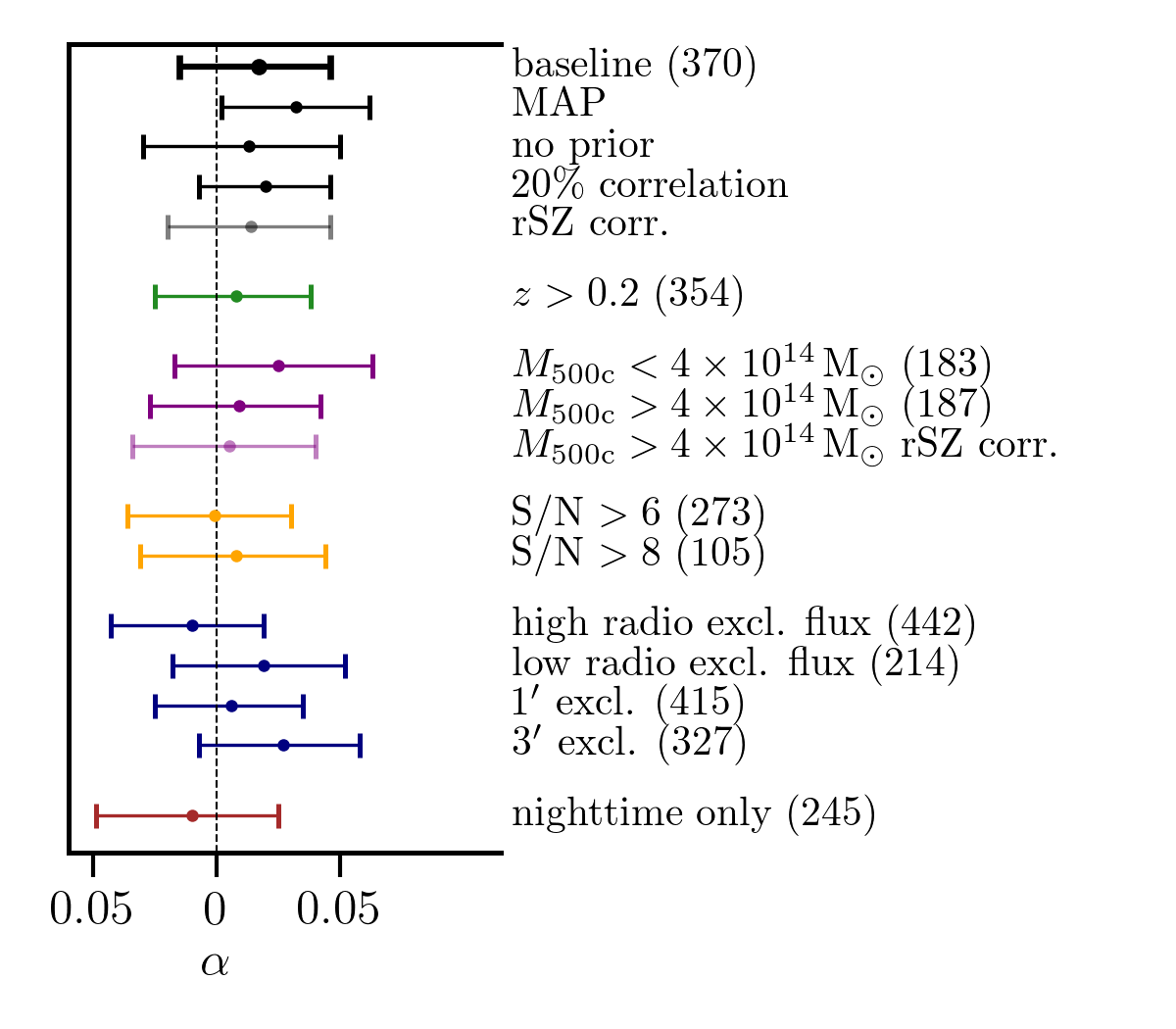}
    \caption{The point estimation for $\alpha$
    (posterior mean, except for the second row that reports MAP) with the associated $68\%$ credible interval derived from the posterior distribution after marginalizing over all nuisance parameters.
    Numbers in parentheses indicate how many clusters were included in each test.
    The top, black point is our baseline result with the selection criteria detailed in the main text. 
    Colored lines denote various systematics checks (see text for details). From top to bottom, these are tests for: the cluster redshift and completeness (green),
    the cluster mass and relativistic correction (purple),
    the S/N selection bias (orange),
    contamination from radio sources (blue)
    and systematics associated with day-time data (brown). 
    }
    \label{fig:splits}
\end{figure}

Figure~\ref{fig:posterior_corner} shows the marginalized posterior distribution for $\alpha$ and $\eta$.
Using the mean of the marginalized posterior distribution, we report the following constraint on the CMB temperature evolution:
\begin{equation}
    \alpha=0.017^{+0.029}_{-0.032}.
 \label{eq:result}
\end{equation}
This is fully consistent with no deviation from the adiabatic evolution.
In addition to the posterior mean value we quote as baseline result, we also report the MAP estimation $\alpha=0.032\pm0.030$ (where the error bar is the 68\% credible interval of the posterior distribution), which is also consistent with the standard temperature evolution model within ${\sim}1\sigma$. 
Since the estimate of $\alpha$ is highly degenerate with the relative calibration $\eta$, the result is sensitive to the assumed prior on $\eta$. 
Lifting this prior constraint we find $\alpha=0.013\pm0.040$ and $\eta=1.004\pm0.026$, consistent with baseline result. 
The comparison of these results are summarized in Figure \ref{fig:splits}.

\begin{figure}
    \centering
    \epsscale{1.4}
    \plotone{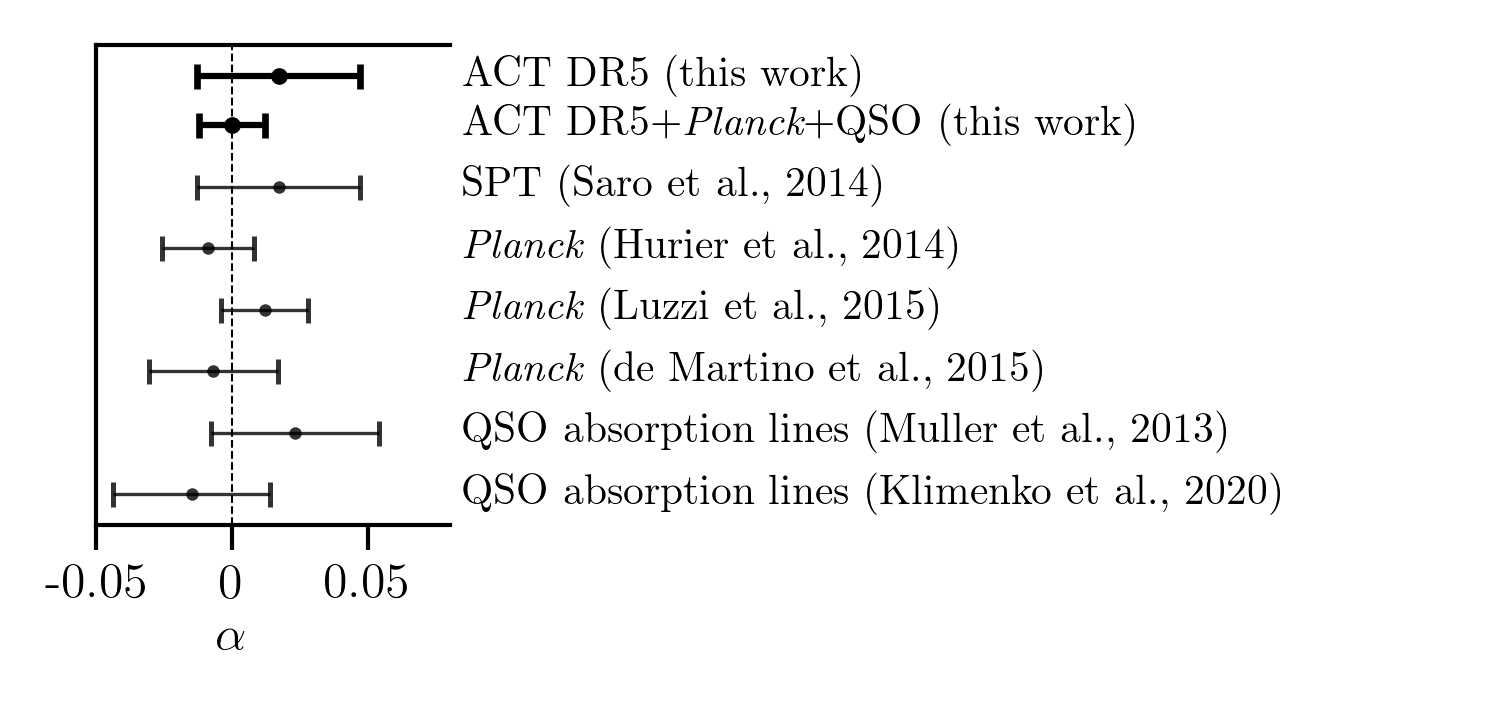}
    \caption{Constraints on $\alpha$ from tSZ measurements and quasar absorption line studies. 
    The result combining ACT, \planck (except for \citealt{Luzzi2015}: see text), and quasar absorption lines \citep{Muller2013, klimenko/etal:2021} is shown in the second row.}
    \label{fig:previous_compare}
\end{figure}

In Figure~\ref{fig:previous_compare} we show our result next to previous results obtained from SZ measurements of clusters from SPT \citep{SPT2014} and \planck \citep{Hurier+2014, Planck2015_martino, Luzzi2015}, and from quasar absorption lines \citep{Muller2013, klimenko/etal:2021}.
The uncertainty in our measurement is comparable to \cite{SPT2014} which used a similar two-frequency observation, despite the improvement in statistical errors from the larger sample with higher S/N threshold. 
This is because our total error is dominated by the two sources of uncertainty in addition to the statistical uncertainties that we account for in our likelihood analysis. 
The first, and biggest, of these comes from the uncertainty of the relative calibration of our two frequency bands, parameterized by $\eta$ in our likelihood \citep[Equation~\ref{eq:likelihood},][]{Luzzi2009}. 
We note that this has not been explicitly marginalised over in some previous studies \citep{SPT2014, Planck2015_martino}. 
The second source of uncertainty comes from covariance with the nuisance parameters $\{\mu_{i,150}\}$ that we marginalize over. This factor, which is not included in the analysis of \citet[][]{SPT2014}, makes up about $10\%$ of our uncertainty budget. 

\begin{figure*}
\epsscale{1.2}
\plotone{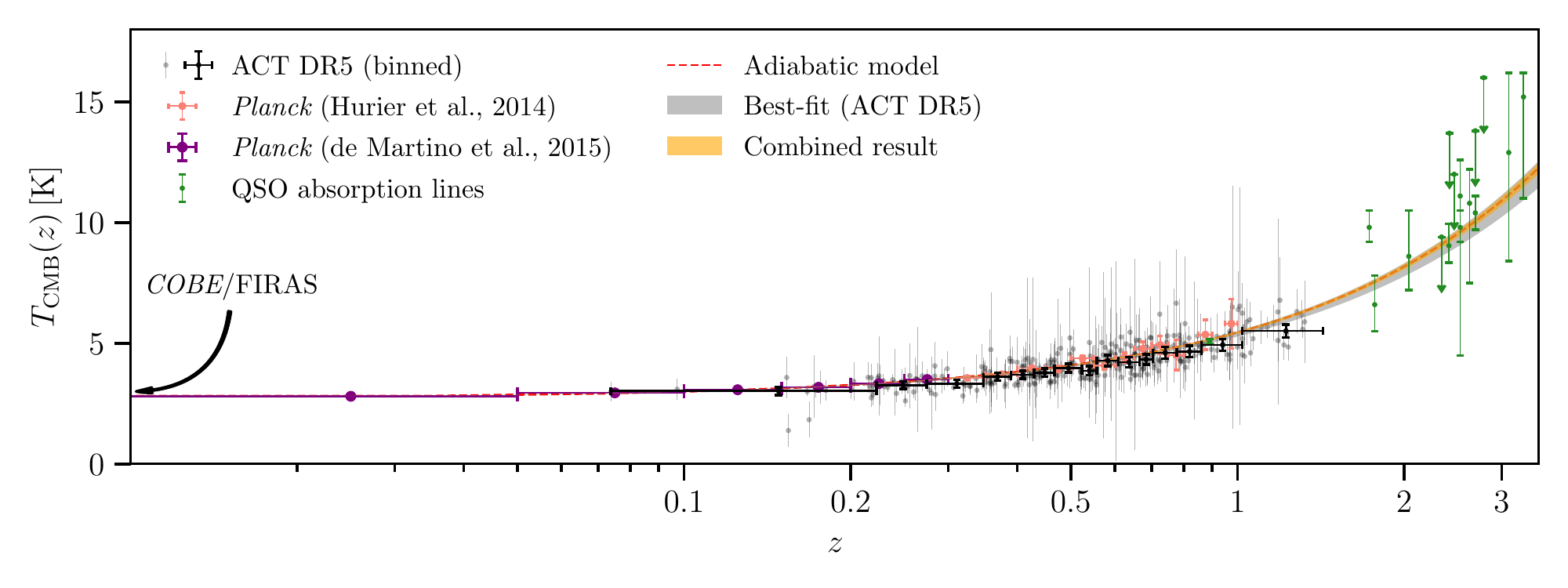}
\caption{CMB temperature estimation inferred from each of the \Ncluster clusters in the sample by inverting the likelihood function Equation \ref{eq:likelihood} (grey bars). 
The error bar of each point is the $68\%$ confidence interval for each $\mathcal{L}_i$.
For visualization purpose, we bin our measurements in redshift and show them as black error bars. 
This binning is arbitrary and is not used for our analysis.
We also include results from two independent analyses of \textit{Planck} clusters,  which estimated the redshift-bin-averaged temperature measurements \citep{Hurier+2014, Planck2015_martino}, and from quasar absorption line studies \citep{Muller2013, klimenko/etal:2021}.
The dashed line marks the standard adiabatic evolution of $T(z)$.
The shaded grey region is our best-fit constraint and is consistent with the adiabatic model to within its uncertainty of $\Delta \alpha =0.03$.
The shaded orange region marks the combined constraints with uncertainty of $\Delta \alpha =0.012$.
\label{fig:T_z}}
\end{figure*}

The constraint on $\alpha$ can be improved by combining the results derived from independent data sets in an inverse variance weighted average.
Our analysis has been done on a sub-sample of 310 clusters above redshift $0.3$ that does not overlap with PSZ1 \citep{PSZ}. 
Combining the $\alpha$ constraint from this sample with those derived using 481 clusters with $z<0.3$ from the X-ray selected \planck sample \citep{Planck2015_martino}, using 267 clusters from a \planck SZ-selected sample above $z=0.3$ \citep{Hurier+2014, Avgoustidis2016},  and using the quasar absorption line studies \citep{Muller2013, klimenko/etal:2021},  we find a joint constraint of $\alpha=-0.001\pm0.012$, assuming independent Gaussian errors in each result.
The individual datasets used for the combination are also shown in Figure \ref{fig:T_z}.
We exclude other results from this combination because they do not indicate their cluster catalog explicitly and we wish to avoid overlap with our sample; the statistical improvement by including them would be marginal.

We performed a battery of tests to examine the impact of other potential systematics not accounted for in our formalism.
A summary of the results from these tests, which we describe below, is shown in Figure \ref{fig:splits}.

\paragraph{Correlated error}
As mentioned in Section \ref{sec:parameter_inference}, correlated astrophysical sources would cause correlated errors in the measurement at two frequencies. 
To test this effect, we modified the likelihood function in Equation \ref{eq:likelihood} by introducing the Pearson Product-Moment correlation factor $\rho=0.2$ for all clusters \citep{JackOS2021}.
The two individual Gaussian functions in the numerator are replaced with the 2D Gaussian function and the complementary error function are replaced with the 2D integrals to account for the adjusted normalization.
The computational cost for evaluating this generalized likelihood function is increased significantly. Therefore, we directly sample the 2D marginalized likelihood of $\{\alpha, \eta\}$ with a Gaussian process emulation technique \citep{Pellejero-Ibanez2020} using the \texttt{GPy} package \citep{gpy2014}.\footnote{http://sheffieldml.github.io/GPy/}
The result is $0.020^{+0.026}_{-0.027}$, consistent with the baseline result and with a 10\% improvement in the constraining power.
Since the simulation pipelines we have used to validate our methods do not implement this novel fitting procedure, we do not use it for our baseline result, but future studies could benefit from using this treatment from end to end.

\paragraph{Completeness}
Simulations with mock cluster catalogs indicate that the completeness of the ACT DR5 cluster sample drops at redshifts below 0.2 \citep{Hilton2020}. 
To check the potential impact from this redshift-dependent selection, we performed the analysis on a subsample of clusters with $z>0.2$, where there is a roughly uniform $90\%$ completeness for cluster mass $M_\mathrm{500c}>3\times10^{14}\,\mathrm{M}_\odot$,\footnote{Similar to $\theta_{500c}$ defined above, $M_\mathrm{500c}$ is the mass enclosed within a radius where the density is more than 500 times the critical density of the Universe. 
We estimate it from the SZ-mass scaling relation as in \citet{Hilton2020}} and obtained a consistent result.

\paragraph{Cluster mass}
We checked the dependence of our result on cluster mass \citep[assuming the mass--scaling relation from ][]{Arnaud_2010} by performing the analysis on subsets of clusters with mass $M_\mathrm{500c}$ above and below $4\times 10^{14}\,\mathrm{M}_\odot$. 
The low-mass sample shows a slightly larger deviation from $\alpha=0$, which could be related to the higher contamination from the infrared emission as the tSZ S/N is typically lower for this sample. 
The massive clusters are expected to have a higher relativistic SZ (rSZ) correction due to their higher virial temperatures, and one might therefore anticipate that the rSZ effect could masquerade as a non-zero $\alpha$.
Using the mass--temperature scaling relation from \cite{Arnaud_2005} and the rSZ model up to the fourth term in temperature from \cite{Itoh1998}, we correct for the rSZ effect for the high-mass and the baseline sample, and find little change in $\alpha$ (light purple and grey points in Figure~\ref{fig:splits}).

\paragraph{Infrared Emission}
Infrared emission from dusty, star-forming galaxies within the clusters could be biasing the tSZ signal measurement.
However, our ability to perform an adequate component separation is limited by the frequency coverage.
Our matched-filter-based photometry helps alleviate this bias as the infrared emission of the dusty galaxies in the cluster is more extended than the tSZ signal \citep{Planck23}.
Additionally, \cite{JackOS2021} found that the ACT cluster sample does not, on average, show evidence for infrared emission after stacking the clusters on ACT 224\,GHz or \textit{Herschel} observations (albeit using a sample drawn from a smaller region than is covered by the DR5 maps; see \citealt{HerschelDR1,HerschelDR2}).
As discussed in Section \ref{sec: websky-sim}, the positive bias in $\alpha$ could be further reduced by raising the S/N threshold for cluster selection. 
We run our fits with several higher S/N thresholds, shown in Figure~\ref{fig:splits}, and find the result insensitive to the choice of threshold.
This demonstrates that the threshold $\xi>5.5$ is sufficient for reducing the infrared emission, in agreement with the results from the \texttt{Websky} simulation. Nevertheless, a sub-$1\sigma$, positive bias due to the CIB could still be present (Section \ref{sec: websky-sim}).

\paragraph{Radio Emission}
Recently \cite{JackOS2021} and Dicker et al., (in prep.) have shown evidence that the radio point sources could be contaminating the tSZ signal measurement. However, it is hard to directly correct for it using low frequency surveys due to the spread in the sources' spectral indexes. 
To guard against the radio emission from cluster-member galaxies biasing tSZ signal measurements, we removed clusters within $2\arcmin$ of radio sources in the joint NVSS and SUMSS catalog with flux density above certain thresholds (see Section ~\ref{ssec:sample_selection}).
To test whether this selection is effective at rendering radio contamination negligible, we raise the flux threshold by a factor of 2 (i.e. relax the exclusion criteria) to select a sample that is potentially more contaminated by radio emissions, and then remove the flux threshold to obtain a sample that is less contaminated.
Due to the ringing of the matched filters, sources at intermediate radii from the center of the cluster could negatively bias the tSZ decrement (for a 2.4$\arcmin$ filter, this negative infill happens between 2$\arcmin$-4$\arcmin$: Dicker et al., in prep.). 
In our case, we use matched filters with a variety of scales, with the majority of cluster photometry extracted using scales within 2$\arcmin$-4$\arcmin$. 
To test the effect of the radio source infill at different distances, we alter the exclusion radius (1$\arcmin$ and 3$\arcmin$) for the source cut while keeping the baseline flux threshold.
The results are shown as blue bars in Figure ~\ref{fig:splits}. 
A ${\sim}0.5\sigma$ shift of $\alpha$ in the negative direction occurs when more clusters with potential radio sources association are included. Nevertheless, the result is consistent within uncertainties, and we note that the best-fit $\alpha$ barely changes when the more aggressive cut is used. 
This indicates that the baseline cut is sufficient for removing radio contamination.

\paragraph{Nighttime-only Data}
Our SZ measurements derive additional precision from daytime data that are included in ACT DR5 maps \citep{Hilton2020}. 
However, time-dependent beam deformations and, to a lesser extent, pointing offsets, both of which are due to the Sun heating the telescope structure, are not fully measured and corrected for in DR5 maps \citep{ACT_DR5_coadd}.
We therefore perform the same analysis on maps made from nighttime-only data as a consistency test.
The result, as shown in Figure \ref{fig:splits}, is consistent with the baseline constraint.
It is worth noting that this split is also correlated with the high-mass splits as low-mass clusters are hardly detected in this noisier map with same S/N threshold $\xi=5.5$.

\section{Conclusions}
Using ACT DR5 data, we have measured the tSZ decrement at 98 and 150~GHz for \Ncluster galaxy clusters in the redshift range $0.07<z<1.4$ using a matched-filter technique. 
We find that the tSZ signal ratio between the two bands is consistent with the CMB temperature evolution in the standard $\Lambda$CDM cosmological model. For a temperature evolution parametrized as \mbox{$T(z) = T_0 (1+z)^{1-\alpha}$} \citep{Lima+2000}, we place a constraint, \mbox{$\alpha=0.017^{+0.029}_{-0.032}$} (or alternatively, \mbox{$\alpha=0.032\pm0.030$} for the MAP value) on the potential deviation from the adiabatic evolution using ACT data alone. This measurement is comparable to the most recent results based on other tSZ data \citep{SPT2014,Hurier+2014,Planck2015_martino,Luzzi2015}.
In obtaining this result, we refined the likelihood analysis from previous studies, and the robustness of this approach has been demonstrated with simulations.
Combining this with previous results from independent data \citep{Planck2015_martino,Hurier+2014, Muller2013, klimenko/etal:2021}, we report an updated composite measurement of \mbox{$\alpha=-0.001\pm0.012$}.

Current constraining power using galaxy clusters observed by ACT is mainly limited by the degeneracy with the relative calibration factor (including the bandpass uncertainties) between two bands. 
This should be improved with wide-frequency-coverage data from future data releases from ACT and the Simons Observatory \citep{SO}. 
Their higher frequency channels will also enable better cleaning of the infrared emissions, a major source of systematic uncertainty in the tSZ photometry at high redshifts.

\begin{acknowledgments}
ACT was supported by the U.S. National Science Foundation through awards AST-0408698, AST-0965625, and AST-1440226 for the ACT project, as well as awards PHY-0355328, PHY-0855887 and PHY-1214379. Funding was also provided by Princeton University, the University of Pennsylvania, and a Canada Foundation for Innovation (CFI) award to UBC. ACT operates in the Parque Astron\'omico Atacama in northern Chile under the auspices of the Chilean National Agency for Research and Development (ANID).  The development of multichroic detectors and lenses was supported by NASA grants NNX13AE56G and NNX14AB58G. Detector research at NIST was supported by the NIST Innovations in Measurement Science program. 

Computations were performed on Hippo at the University of KwaZulu-Natal; data products used in this work also relied on computations on Cori at NERSC as part of the CMB Community allocation, on the Niagara supercomputer at the SciNet HPC Consortium, and on Feynman and Tiger at Princeton Research Computing. SciNet is funded by the CFI under the auspices of Compute Canada, the Government of Ontario, the Ontario Research Fund--Research Excellence, and the University of Toronto.

ADH is grateful for support from the Sutton Family Chair in Science, Christianity and Cultures.
RH acknowledges funding from the NSERC Discovery Grants program, CIFAR Azrieli Global Scholars program and the Alfred P. Sloan Foundation.
KM and MH acknowledge support from the National Research Foundation of South Africa (grant number 112132).
NS acknowledges support from NSF (grant number AST-1907657). CS acknowledges support from the Agencia Nacional de Investigaci\'on y Desarrollo (ANID) under FONDECYT (grant number\ 11191125).
ZX is supported by the Gordon and Betty Moore Foundation.
\end{acknowledgments}

\software{Astropy \citep{Astropy}, NumPy \citep{Numpy}, SciPy \citep{Scipy} Matplotlib \citep{Matplotlib}, Cython \citep{Cython}, emcee \citep{emcee}, GPy\citep{gpy2014}}.

\bibliography{main}
\end{document}